# Laser-Frequency Stabilization via a Quasimonolithic Mach-Zehnder Interferometer with Arms of Unequal Length and Balanced dc Readout

Oliver Gerberding,[1,*] Katharina-Sophie Isleif,[2,†] Moritz Mehmet,[1,2] Karsten Danzmann,[1,2] and Gerhard Heinzel[1]

[1]*Albert Einstein Institute, Max Planck Institute for Gravitational Physics,
Callinstrasse 38, 30167 Hannover, Germany*

[2]*Institute for Gravitational Physics, Leibniz Universität Hannover,
Callinstrasse 38, 30167 Hannover, Germany*

Low-frequency high-precision laser interferometry is subject to excess laser-frequency-noise coupling via arm-length differences which is commonly mitigated by locking the frequency to a stable reference system. This approach is crucial to achieve picometer-level sensitivities in the 0.1-mHz to 1-Hz regime, where laser-frequency noise is usually high and couples into the measurement phase via arm-length mismatches in the interferometers. Here we describe the results achieved by frequency stabilizing an external cavity diode laser to a quasimonolithic unequal arm-length Mach-Zehnder interferometer readout at midfringe via balanced detection. We find this stabilization scheme to be an elegant solution combining a minimal number of optical components, no additional laser modulations, and relatively low-frequency-noise levels. The Mach-Zehnder interferometer is designed and constructed to minimize the influence of thermal couplings and to reduce undesired stray light using the optical simulation tool IFOCAD. We achieve frequency-noise levels below 100 Hz/$\sqrt{\text{Hz}}$ at 1 Hz and are able to demonstrate the LISA frequency prestabilization requirement of 300 Hz/$\sqrt{\text{Hz}}$ down to frequencies of 100 mHz by beating the stabilized laser with an iodine-locked reference.

## I. INTRODUCTION

Laser interferometry is the tool of choice for performing ultraprecise displacement and tilt measurements. One prominent subset of experiments is to apply such displacement measurements to free-floating test masses to investigate the effects of small forces and gravity itself. The most important recent achievements in this regard are certainly the direct detection of gravitational waves by the LIGO detectors [1] as well as the unprecedented low force noise levels demonstrated in the LISA Pathfinder satellite mission [2], a precursor for the space-based gravitational-wave detector LISA [3]. The high displacement and angular sensitivities of laser interferometry have also motivated the implementation of the laser-ranging instrument in the GRACE Follow-On mission to further improve the intersatellite ranging for the determination of Earth's time-varying geoid [4]. Laser interferometry for measurements in the 1-Hz to below 1-mHz regime, as used in these missions, is often limited by laser-frequency noise, which is typically increasing towards lower frequencies and couples into phase measurements via unequal arm lengths.

Laser-frequency stabilization at low frequencies has been extensively researched for the above-mentioned space missions [5], as well as for optical clocks [6] and other metrology experiments [7]. In comparison to such rather demanding stability requirements, the laser-frequency-noise levels required for achieving 1 pm/$\sqrt{\text{Hz}}$ displacement sensing levels for local interferometers in satellites, prominently test mass readouts, are moderate, because they are driven by much shorter arm-length differences. Because of the use of time-delay interferometry [8], these moderate levels are also sufficient as prestabilization levels for the LISA mission, which, even though it is expected to have the longest interferometer arm lengths and arm-length differences of any laser interferometer so far, has been proposed to be operated with frequency stabilities in the order of 300 Hz/$\sqrt{\text{Hz}}$ (assuming an absolute arm-length-ranging accuracy of about 1 m) at low frequencies [9]. The use of ultrastable optical benches in a thermally stable environment on the LISA Pathfinder, LISA, and similar satellite missions has led to experiments using this displacement stability to implement laser stabilizations with intentional unequal arm-length interferometers as frequency sensors. Such techniques have been implemented and tested using kilohertz heterodyne interferometry [10,11], they have been proposed and analyzed for megahertz heterodyne interferometry [9], and optical cavities interrogated using a heterodyne readout technique have also been used [12].

---

[*]contact@olivergerberding.com
[†]katharina-sophie.isleif@aei.mpg.de

Here we implement a related type of laser-frequency-stabilization scheme that uses an unequal arm-length Mach-Zehnder interferometer (MZI) with balanced homodyne dc readout constructed on a glass ceramic with a low thermal expansion coefficient. The main motivation to implement such a homodyne stabilization is based on recent developments and renewed interest in interferometry techniques that rely on some form of self-homodyning [13–16], all of which are in some form compatible with this stabilization scheme. Additionally, the balanced dc readout belongs to a class of straightforward techniques that do not require any type of modulations or ac readout electronics, and, thus, it might be an interesting option also for other experiments requiring some form of frequency-noise reduction.

## II. MACH-ZEHNDER DESIGN AND IMPLEMENTATION

The interferometer layout is designed using the C++-based optical simulation library IFOCAD aiming for an arm-length difference of about 12.5 cm. Numerical optimizations are used to minimize the influence of spurious beams generated by residual reflections at secondary surfaces. To this end, the interferometer design includes two wedged beam splitters enabling us to decouple secondary reflections by changing their propagation angles. A two-dimensional interferometer layout generated by the software is shown in Fig. 1. One additional feature of the interferometer is the third beam splitter placed behind the interference beam splitter to enable diagnostic so-called optical zero measurements.

The interferometer baseplate is made from CLEARCERAM HS, a glass ceramic with a coefficient of thermal expansion of $1 \times 10^{-8}$ 1/K at ambient temperature and with dimensions of $13.5 \times 13.5$ cm$^2$ and a thickness of 3.6 cm. The light is brought onto the bench using a commercial fiber collimator creating a collimated beam with a 1-mm waist radius mounted in an adjustable tip-tilt, $x$-$y$ mount. The mount itself is mounted on an adapter made from Invar that is glued using an epoxy to the side of the baseplate. The components are coated fused-silica parts with a thickness of 7 mm.

The component alignment is done using a coordinate measurement machine and a combination of template-assisted positioning for uncritical components and an adjustable pointing finger assembly [17] for the recombination beam splitter. For the alignments of the input beam relative to the template, we use an in-house-developed beam measurement technique [18]. The bonding of the components is done using an UV-cured optical adhesive that is applied using only minimal amounts of glue to achieve thin, planar bonding layers. Similar techniques have been implemented previously, for example, based on a two-component epoxy [19]. The final recombination beam-splitter alignment is done with a continuous contrast monitoring by applying a deep frequency modulation [14,15], enabling straightforward optimization with the adhesive already applied to the component before UV curing. Figure 2 shows a photograph of this final assembly stage achieving an optical contrast of more than 95% that remains constant during the UV curing, which takes less than 2 min. No degradation of contrast is observed after four months of storage and operations, indicating also a decent long-term stability of the UV bond.

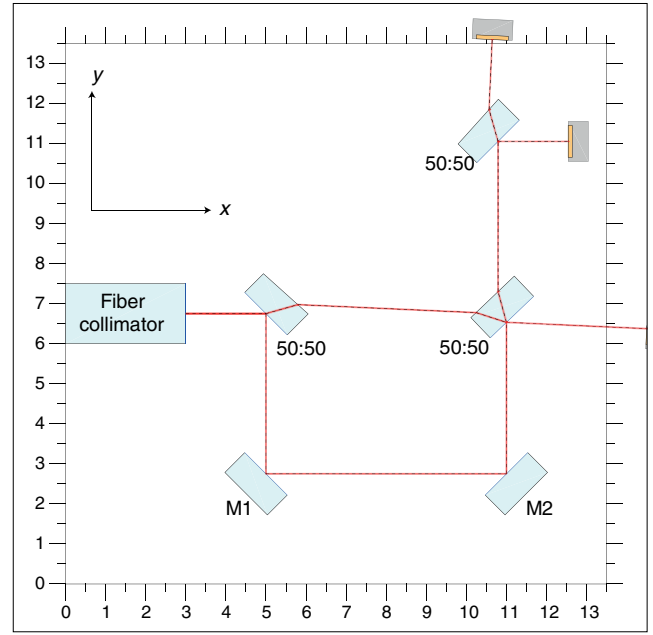

FIG. 1. Layout of the Mach-Zehnder interferometer with additional output splitting in the north output of the recombination beam splitter created using a 2D output of the IFOCAD C++ library.

## III. FREQUENCY-STABILIZATION SETUP

To probe the displacement stability of the interferometer, we place it into a thermally isolated vacuum chamber and set up a balanced dc readout scheme depicted in Fig. 3. To achieve a balanced operation, we employ a second externally placed beam splitter in the east output and use its reflectivity dependence on the macroscopic beam incidence angle to achieve matched power levels on both photodiodes at the midfringe operation point. A low-noise, low-drift operational amplifier in combination with ultrastable photodiode bias voltages is implemented to perform a direct current subtraction followed by a high-gain transimpedance amplification. The so-generated sinusoidal output signal contains regular zero crossings that can be used for locking

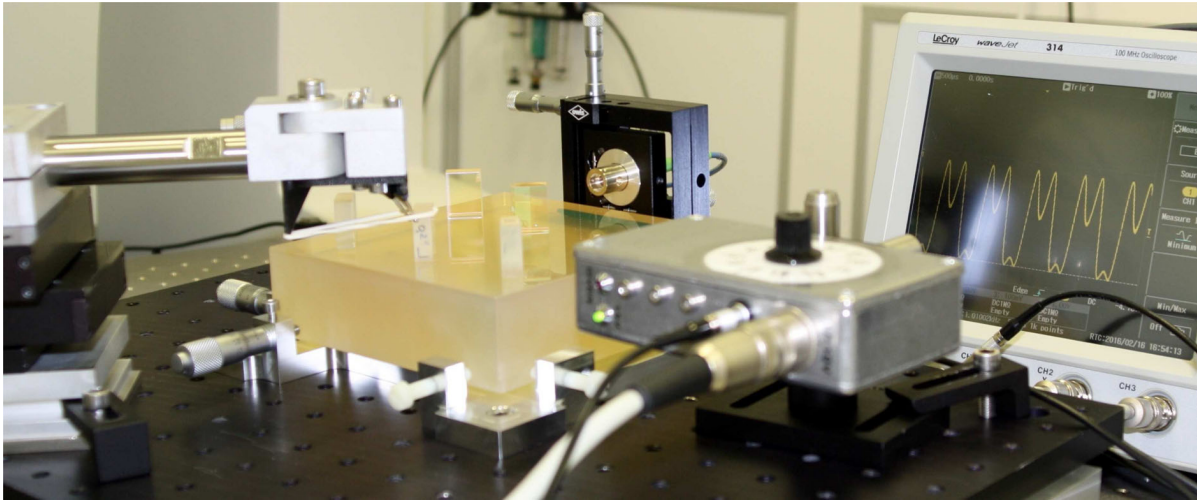

FIG. 2. Photograph of the construction of the Mach-Zehnder interferometer. Shown is the alignment stage of the recombination beam splitter using a pointing finger assembly while monitoring the interferometric contrast using a strong laser-frequency modulation.

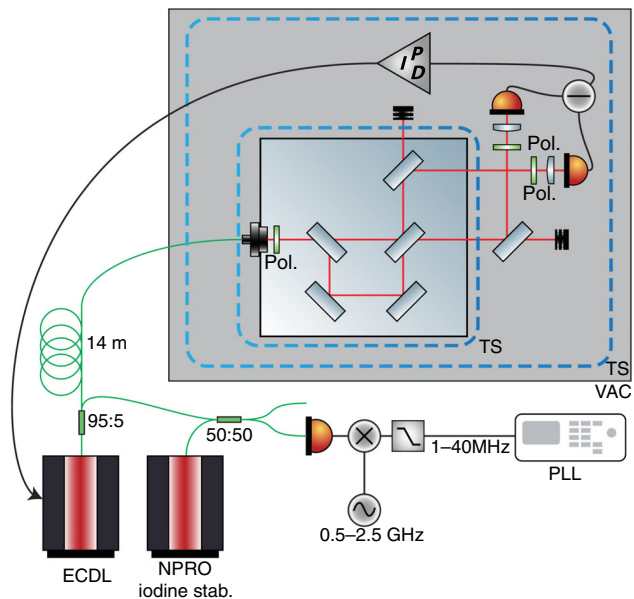

FIG. 3. Sketch of the frequency-stabilization experiment. The optical setup outside the vacuum chamber (VAC) consists of single-mode, polarization-maintaining fiber components. The beat frequency is first mixed down to a frequency between 1 MHz and 40 MHz and then tracked by a digital phase-locked loop (PLL) implemented in a LISA-like phase meter. The vacuum level during the experiment is on the order of $5 \times 10^{-6}$ mbar. The interferometer and the analogue electronics are placed inside a thermal shield (TS) mounted in the vacuum chamber via thermally isolating feet. An additional thermal shield is placed around the baseplate of the interferometer and extended to cover the fiber coupler with multilayer insulation foil. The heat generated by the electronics is understood to be the main driver of the observed long-term drifts, with a thermal equilibrium still not achieved after a week of pumping. The optical input power of the interferometer is on the order of 10 mW. A dedicated analogue controller (PID) is used to provide feedback to the ECDL.

without the need for subtracting an additional reference signal.

The so-generated voltage is then used as input to a frequency-stabilization control loop that feeds back to a rapidly tunable external cavity diode laser (ECDL, TLB-6821, New Focus [20]) situated outside the vacuum chamber connected to the interferometer via a 14-m-long fiber. Around 5% of the light is picked off and fed into a fiber beam splitter, interfering it with light from an iodine-stabilized nonplanar ring oscillator (NPRO) laser [21] that is used as frequency reference. Using a tunable demodulation scheme, we measure the beat-note frequency stability using a digital phase measurement system [22]. Given an arm-length difference of $\Delta l = 12.5$ cm of the MZI, which we confirm using a measurement of the free spectral range of the interferometer, we can use the measured frequency fluctuations $\delta f$ to determine the effective displacement noise $\delta l$ using the laser wavelength $\lambda_0 \approx 1064.5$ nm and the speed of light $c$,

$$\delta l = \Delta l \times \frac{\delta f}{f_0} = \Delta l \times \frac{\delta f \lambda_0}{c}. \quad (1)$$

By reversing this calculation, we can also calculate the frequency-noise level that is required to achieve a displacement noise of 1 pm/$\sqrt{\text{Hz}}$, the standard goal for the local interferometry in LISA and LISA Pathfinder. This level is 4.5 kHz/$\sqrt{\text{Hz}}$, taking into account an additional factor of 2 in the displacement sensing that is gained by using a reflection setup. If it is proven that the Mach-Zehnder interferometer can provide this level of stability, it will be a suitable frequency reference for interferometer setups that aim for the same level of displacement noise and use equivalent or shorter arm-length differences.

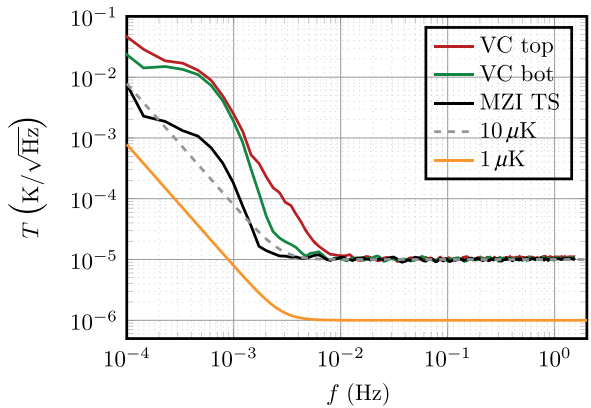

FIG. 4. Temperature spectral densities of the vacuum chamber (VC) and the MZI during the performance measurements. Shown are also the thermal stability (TS) goal for the LISA optical bench environment, a $1~\mu\text{K}/\sqrt{\text{Hz}}$ white noise relaxed towards lower frequencies, as well as a 10-time increase level for comparisons.

To probe the full performance level of this stabilization scheme, we immediately implement a number of known techniques that have been crucial for achieving the performance in LISA breadboarding experiments [10]. This includes prominently the placement of thin-film polarizers with ultrahigh extinction ratio after the fiber coupler and in front of both photodiodes to reduce the influence of parasitic interferences due to residual beams in the undesired, orthogonal polarization. Additionally, we use focusing lenses in front of the photodiodes to reduce beam-walk effects, and we thermally isolate the long fibers routed outside the vacuum chamber using passive means. The experiment is placed inside a thermal shield mounted via thermally resistive materials in a vacuum chamber with an additional passive thermal isolation layer on the outside. The interferometer itself is covered by a second thermal shield with an attached temperature sensor, and the fiber collimator and its mount are additionally covered with multilayer insulation foil. The achieved temperature stabilities are measured, and the corresponding linear spectral densities computed after subtracting a linear drift are shown in Fig. 4.

## IV. RESULTS AND DISCUSSION

The best frequency-noise stability achieved with the current setup is shown in Fig. 5 in comparison to the free-running noise of the external cavity diode laser and the noise between two iodine-stabilized systems that had been measured previously indicating the fundamental performance limit of the setup. At 1 Hz, we reach this limit with $90~\text{Hz}/\sqrt{\text{Hz}}$, which corresponds to a displacement noise of $40~\text{fm}/\sqrt{\text{Hz}}$. The $4.5~\text{kHz}/\sqrt{\text{Hz}}$ sensitivity (equivalent to $1~\text{pm}/\sqrt{\text{Hz}}$) is achieved at all frequencies above 5 mHz, with a noise feature exceeding this level at lower frequencies. Some couplings due to vibrations and acoustics are visible at frequencies above 1 Hz, and it is unknown whether this is introduced by the interferometer sensing or by excess fluctuations induced in the tunable diode laser that are not sufficiently suppressed by the loop gain. The unity gain frequency of the stabilization is in the order of 1 kHz and mainly limited by the bandwidth of the frequency actuation piezo amplifier.

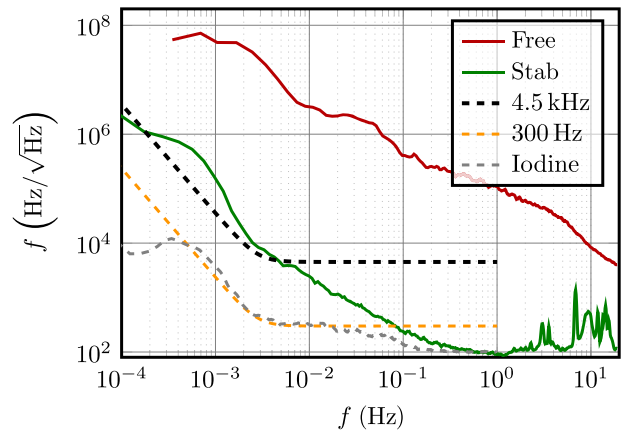

FIG. 5. Frequency spectral densities of the beat note between the iodine-stabilized NPRO laser and the ECDL, once free running and once stabilised to the MZI. Shown are also a frequency-noise level of $4.5~\text{kHz}/\sqrt{\text{Hz}}$ and the LISA frequency prestabilization requirement of $300~\text{Hz}/\sqrt{\text{Hz}}$. The graph also shows the frequency noise measured between two iodine-stabilized NPRO lasers, which is a measure for the lowest noise levels achievable with an iodine-stabilized frequency reference [23]. The frequency-readout-noise floor (not shown) is measured separately using only electronic signals and is on the order of $1~\text{Hz}/\sqrt{\text{Hz}}$, slightly increasing towards lower frequencies.

The excess noise observed at around 0.3 mHz is identified as outside excitation of our thermal environment, being clearly visible also in temperature sensors placed at the top and bottom of the inside of the vacuum chamber, as shown in Fig. 4. Subtracting linear drifts from both the temperature measurement on top of our interferometer as well as the beat-frequency measurement, we find a correlated time series shown in Fig. 6 with a coupling factor of about 0.5 MHz/1 mK. The predicted coupling based on the coefficient of thermal expansion of the base material and the arm-length difference is significantly below the observed value. Based on our IFOCAD simulations, we assume that temperature fluctuations couple into the longitudinal phase measurement mainly via beam tilts. This coupling is dominant in the interferometer due to the wedged components and the comparably high thermal coefficients of the commercial fiber collimator and its adjustable mount, as measured by Dehne *et al.* [10,24] for such assemblies.

Even with this nonideal input beam configuration, we achieve the LISA frequency prestabilization requirement down to 100 mHz. Extending this performance down to 1-mHz frequencies requires the inclusion of more stable monolithic fiber collimators [25] or further reduction of the low-frequency temperature fluctuations. A comparison of the measured levels with the thermal stabilities expected for LISA is shown in Fig. 4, and it is, however, quite encouraging that these levels can even be achieved with the current device.

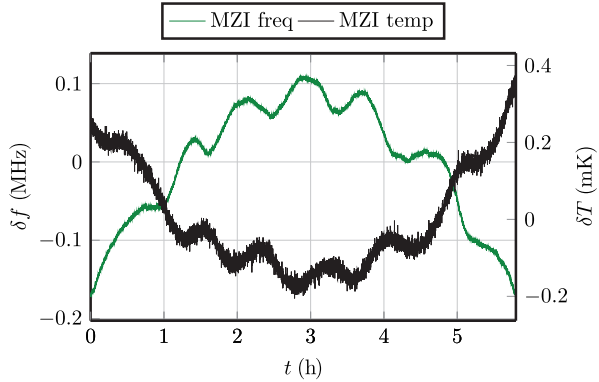

FIG. 6. Shown are the deviation of the measured beat frequency and the temperature measured at the thermal shield on top of the interferometer. Linear drifts of 673.13 Hz/s and 0.96 $\mu$K/s, respectively, are subtracted, as well as a mean value of the temperature of 25.669 65°.

Based on the measured frequency-noise levels, we regard the unequal arm-length Mach-Zehnder interferometer as a prime frequency reference candidate for interferometry techniques that rely on self-homodyning and unequal arm lengths. The achievable stabilization levels for such experiments depend on the described limitations of the interferometer and on the specific phase readout algorithms, which might be limited by additional effects that are not described here.

A number of further improvements can be made to lower noise sources which are currently not identified to be limiting. These include the replacement of the Si photodiodes with devices based on (In,Ga)As, largely increasing the photodiode responsivity and signal currents, hence, reducing shot noise and electronic noise coupling. The addition of a local intensity stabilization also improves the robustness of the system against power fluctuations of the light provided to the interferometer. The current usage of additional beam splitters in the output ports does half the usable optical power but can also enable the implementation of redundant photodiodes, an alternative that might be considered for space applications.

## ACKNOWLEDGMENTS

The authors thank Daniel Penkert and Daniel Schütze for their help with the component positioning and beam measurement used for the interferometer construction. We thank Maike Lieser for supplying us with the reference measurement of the iodine-stabilized laser-frequency noise. We also thank Stefan Ast for the help with the vacuum setup. The authors thank the DFG Sonderforschungsbereich 1128 Relativistic Geodesy and Gravimetry with Quantum Sensors, geo-Q, for financial support. We also acknowledge support by the Deutsches Zentrum für Luft- und Raumfahrt with funding from the Bundesministerium für Wirtschaft und Technologie (Project Reference No. 50 OQ 0601).

[1] B. P. Abbott et al. (LIGO Scientific Collaboration and Virgo Collaboration), Observation of Gravitational Waves from a Binary Black Hole Merger, Phys. Rev. Lett. 116, 061102 (2016).

[2] M. Armano et al., Sub-Femto-g Free Fall for Space-Based Gravitational Wave Observatories: LISA Pathfinder Results, Phys. Rev. Lett. 116, 231101 (2016).

[3] K. Danzmann et al. (eLISA Consortium), The Gravitational Universe: Whitepaper for the ESA L2/L3 selection, arXiv:1305.5720.

[4] B. S. Sheard, G. Heinzel, K. Danzmann, D. A. Shaddock, W. M. Klipstein, and W. M. Folkner, Intersatellite laser ranging instrument for the GRACE follow-on mission, J. Geodes. 86, 1083 (2012).

[5] W. M. Folkner, G. de Vine, W. M. Kleipstein, K. McKenzie, R. E. Spero, R. Thompson, N. Yu, M. Stephens, J. Leitch, R. Pierce, T. T. -Y. Lam, and D. A. Shaddock, "Laser frequency stabilization for GRACE-2", in Proceedings of Earth Science Technology Forum, 2011, https://esto.nasa.gov/conferences/estf2011/author.html.

[6] T. Kessler, C. Hagemann, C. Grebing, T. Legero, U. Sterr, F. Riehle, M. Martin, L. Chen, and J. Ye, A sub-40-mHz-linewidth laser based on a silicon single-crystal optical cavity, Nat. Photonics 6, 687 (2012).

[7] T. Schuldt, K. Dringshoff, A. Milke, J. Sanjuan, M. Gohlke, E. V. Kovalchuk, N. Grlebeck, A. Peters, and C. Braxmaier, High-performance optical frequency references for space, J. Phys. Conf. Ser. 723, 012047 (2016).

[8] M. Tinto and S. V. Dhurandhar, Time-delay interferometry, Living Rev. Relativ. 8, 1 (2005).

[9] B. S. Sheard, G. Heinzel, and K. Danzmann, LISA long-arm interferometry: An alternative frequency pre-stabilization system, Classical Quantum Gravity 27, 084011 (2010).

[10] M. Dehne, M. Tröbs, G. Heinzel, and K. Danzmann, Verification of polarising optics for the LISA optical bench, Opt. Express 20, 27273 (2012).

[11] G. Heinzel, V. Wand, A. Garca, O. Jennrich, C. Braxmaier, D. Robertson, K. Middleton, D. Hoyland, A. Rdiger, R. Schilling, U. Johann, and K. Danzmann, The LTP interferometer and phasemeter, Classical Quantum Gravity 21, S581 (2004).

[12] J. Eichholz, D. B. Tanner, and G. Mueller, Heterodyne laser frequency stabilization for long baseline optical interferometry in space-based gravitational wave detectors, Phys. Rev. D 92, 022004 (2015).

[13] A. Sutton, O. Gerberding, G. Heinzel, and D. Shaddock, Digitally enhanced homodyne interferometry, Opt. Express 20, 22195 (2012).

[14] O. Gerberding, Deep frequency modulation interferometry, Opt. Express 23, 14753 (2015).

[15] K.-S. Isleif, O. Gerberding, T. S. Schwarze, M. Mehmet, G. Heinzel, and F. G. Cervantes, Experimental demonstration of deep frequency modulation interferometry, Opt. Express **24,** 1676 (2016).

[16] T. Kissinger, T. O. Charrett, and R. P. Tatam, Range-resolved interferometric signal processing using sinusoidal optical frequency modulation, Opt. Express **23,** 9415 (2015).

[17] D. Penkert, Diploma thesis, school Leibniz Universität Hannover, 2016.

[18] D. Schütze, V. Müller, and G. Heinzel, Precision absolute measurement and alignment of laser beam direction and position, Appl. Opt. **53,** 6503 (2014).

[19] T. Schuldt, M. Gohlke, D. Weise, U. Johann, A. Peters, and C. Braxmaier, Picometer and nanoradian optical heterodyne interferometry for translation and tilt metrology of the LISA gravitational reference sensor, Classical Quantum Gravity **26,** 085008 (2009).

[20] New Focus TLB-6800 Data Sheet No. DS-051202(07/14), 2014.

[21] Cohenrent Ultra-Narrow Linewidth CW DPSS Laser System Data Sheet No. MC-015-13-1M0113, 2013.

[22] O. Gerberding, C. Diekmann, J. Kullmann, M. Tröbs, I. Bykov, S. Barke, N. C. Brause, J. J. Esteban Delgado, T. S. Schwarze, J. Reiche, K. Danzmann, T. Rasmussen, T. V. Hansen, A. Enggaard, S. M. Pedersen, O. Jennrich, M. Suess, Z. Sodnik, and G. Heinzel, Readout for intersatellite laser interferometry: Measuring low frequency phase fluctuations of high-frequency signals with microradian precision, Rev. Sci. Instrum. **86,** 074501 (2015).

[23] M. Lieser, Ph.D. thesis, Leibniz Universität Hannover, 2017.

[24] M. Dehne, Ph.D. thesis, Leibniz Universität Hannover, 2012.

[25] C. J. Killow, E. D. Fitzsimons, M. Perreur-Lloyd, D. I. Robertson, H. Ward, and J. Bogenstahl, Optical fiber couplers for precision spaceborne metrology, Appl. Opt. **55,** 2724 (2016).